\documentclass[12pt]{iopart}
\usepackage{graphicx}
\usepackage{amssymb}
\usepackage{epstopdf}
\usepackage{cite}

\def\thesection{\arabic{section}}

\begin{document}
\DeclareGraphicsExtensions{.pdf,.gif,.jpq}

\title[Model of crystal orientation in nacre]{Theoretical characterization of a model of aragonite crystal orientation in 
red abalone nacre}
\author{S.N. Coppersmith, P.U.P.A. Gilbert, and R.A. Metzler}

\address{
Department of Physics, University of Wisconsin, Madison, WI 53706}
\date{\today}
%\maketitle
%Short title:  Model of crystal orientation in nacre

%PACS:  87.85.J-, 89.75.-k, 87.23.Cc

\begin{abstract}
Nacre, commonly known as mother-of-pearl,
 is a remarkable biomineral that in red abalone consists of 
 layers of 400-nm thick aragonite crystalline tablets confined by organic matrix sheets, with
the $(001)$
 crystal axes of the aragonite tablets oriented to within $\pm 12^o$ from the
normal to the layer planes.
Recent experiments demonstrate that this orientational order 
develops over a distance of tens of layers from
the prismatic boundary at which nacre formation begins.

Our previous simulations of a model in which the order develops because of differential
tablet growth rates (oriented tablets growing faster than misoriented ones)
yield patterns of tablets that agree
qualitatively and quantitatively with the experimental measurements.
This paper presents an analytical treatment of this model, 
focusing on how the dynamical development and eventual degree of
order depend on model parameters.
Dynamical equations for the probability
distributions governing tablet orientations are
introduced whose form can be determined
from symmetry considerations and for which substantial
analytic progress can be made.
Numerical simulations are performed to relate the parameters used
in the analytic theory to those in the microscopic growth model.
The analytic theory demonstrates that the dynamical mechanism is
able to achieve a much higher degree of order than naive estimates
would indicate.
\end{abstract}

\section{Introduction}
\label{sec:introduction}
Nacre, or mother-of-pearl, is a biomineral that attracts the attention of
materials scientists, biologists, and mineralogists as well as physicists because
of its remarkable mechanical properties and its incompletely elucidated formation 
mechanisms~\cite{Mannbook2001,Nud2006,Wei2006,Nas2005,Rou2005}.
Aragonite, a hard but brittle orthorhombic $CaCO_3$ polymorph, accounts
for 95\% of nacre's mass, yet nacre is 3000 times tougher than
aragonite~\cite{Cur77}.
No synthetic composites outperform their components by such large factors.
It is therefore of great interest to understand the mechanisms governing
nacre formation.

Nacre is a layered composite in which organic matrix (OM) sheets alternate with aragonite layers, each of which
consists of tablets of irregular polygonal shape that
 completely fill the space between preformed OM layers.
 In red abalone, the OM layers are 30-nm thick, and the aragonite tablets
 are of thickness 400-500nm and width 5-6 microns~\cite{Met2007,Gil2008}.
The aragonite tablets are crystalline and oriented with their (001) crystal axes
within $\pm 12 ^\circ$ from the normal to the layer plane~\cite{Lev2001,Gil2008}.
TEM and AFM measurements have shown that the OM has pores through
which aragonite can grow~\cite{Sch97,Son2003},
and x-ray and x-ray photoelectron emission spectromicroscopy (X-PEEM)
 measurements demonstrate that nacre has
stacks of co-oriented tablets~\cite{DiM2004,Met2007,Gil2008}, consistent
with the hypothesis that aragonite crystals grow through pores in the OM sheets.

Growth through pores explains how crystal orientation of the aragonite
tablets is maintained between layers in the material, but it does not
necessarily explain the physical mechanism giving rise
to the orientational alignment in the first place.
It is common belief in the biomineralization field that
the alignment of the aragonite crystal c-axes is
due to microscopic chemical templation by the OM~\cite{Add85,Mannbook2001},
with organic molecules providing surfaces that promote aragonite
nucleation with preferred orientations.
However, it is not obvious why such a mechanism would lead to
a high degree of orientational order of the c-axes but not
of the a or b axes of the aragonite tablets.
Moreover,
Ref.~\cite{Gil2008} reports x-ray photoelectron emission spectromicrosopy
(X-PEEM) and microbeam X-ray diffraction that probe the degree of orientational alignment
of the aragonite c-axes~\cite{Met2007,Met2008} and
demonstrate that the orientational alignment of the aragonite tablets 
increases systematically over a length scale of tens of microns
starting from the prismatic boundary at which
nacre growth originates.
As discussed in Ref.~\cite{Gil2008}, this observation suggests
that the aragonite crystal orientation is the result of
a dynamical process.

In Refs.~\cite{Met2007b,Gil2008} a dynamical model is proposed for the
development of orientational
order of the aragonite tablets in nacre in which the presence of co-oriented stacks
of tablets plays an essential role, and the ordering arises
because oriented tablets grow faster than misoriented ones.
Refs.~\cite{Met2007b,Gil2008}
present numerical simulations of the model that successfully reproduce
 several different aspects of
the pattern of tablet orientations in red abalone nacre
measured using X-PEEM.
This paper presents
a more detailed investigation of the model and its
behavior, including closed-form analytic results that apply in the limit that the
probability of nucleating a tablet with
orientation different from the one immediately below is small,
which, based on comparison to the X-PEEM results, is the physically
relevant regime.
We find that the dynamical ordering mechanism can lead to a remarkably high
 degree of orientation of the tablet c-axes, because the width of the
 distribution of tablet orientations decreases very strongly as the
 fraction of tablets that nucleate with the ``wrong'' orientation decreases. 

The analytic characterization of the model uses
methods similar to those used to study mutation-selection
models in population biology~\cite{Kin78,Tur84,Cop99}.
However, the analytic formulation involves some
parameters whose relationship to those
of the growth model is not determined in Ref.~\cite{Gil2008}.
These parameters are examined here, and
it is shown that this relationship is not trivial.
The relationship between the parameter sets
depends on the spatial arrangement of the nucleation
sites, and some aspects are quite insensitive to
changes in the values of the microscopic parameters.
Numerical simulations  and mean-field arguments are used to
relate the parameters in the analytic probabilistic model
to the parameters in the original growth model
for the case when the nucleation sites are chosen randomly
with uniform probability on each layer.

The paper is organized as follows. 
Section~\ref{sec:model} presents the model, while
section~\ref{sec:analytic_theory} presents the analysis of
the probabilistic model that enables one to understand 
qualitatively some features of the behavior, using
 methods that have been developed to study
models relevant to population 
biology~\cite{Kim65,Cro70,Kin78,%
Tur84,Bur96,Cop99,Wax98}.
The fixed point behavior is discussed in subsection~\ref{subsec:fixedpoint},
while the dynamical evolution is discussed in
subsection~\ref{subsec:dynamics}.
Section~\ref{sec:parameters} discusses the relationships between the
parameters of the growth model to those of the model for the
probability distribution for tablet orientations.
Sec.~\ref{sec:discussion} is a discussion, and
the conclusions are presented in Sec.~\ref{sec:conclusions}.
Appendix A presents additional details of the arguments justifying the functional
forms of the equations used in the main text.

\section{The model}
\label{sec:model}
Figure~1 illustrates the basic mechanism of 
nacre growth~\cite{Fri94,Sch97,Mannbook2001}.
First, organic matrix sheets spaced by approximately 0.4 microns are created.
Crystalline aragonite tablets then nucleate on the first layer at
uncorrelated random locations and grow
while confined by the organic matrix sheets.
The crystals in this layer continue to grow and fill out the space in the
layer~\cite{Lowenstambook89,Mannbook2001,Wei81}.
Each nacre tablet in a given layer nucleates and then grows until it reaches
confluence with a neighboring tablets, so that
the resulting tablet pattern resembles a Voronoi construction~\cite{Nud2006,Rou2005},
with tablet in-plane width of order 5 microns.
The crystal orientation of each tablet is highly probable to be the
same as that of the tablet directly below its
nucleation site~\cite{Sch97,Son2003,Car2007},
which reflects the presence of pores
in the organic matrix 
 (typically with diameter $\sim5-50~{\rm nm}$)
 through which aragonite crystals can grow.
 As discussed in~\cite{Nud2006,Gil2008}, there is one nucleation site per tablet,
 with an identifiable structure in the organic matrix that is large enough to have
one or more pores going through it.
%\begin{figure}[htbp]
\begin{figure}[ht]
\vspace{0.5cm}
\hspace{1cm}
\includegraphics[height=6cm]{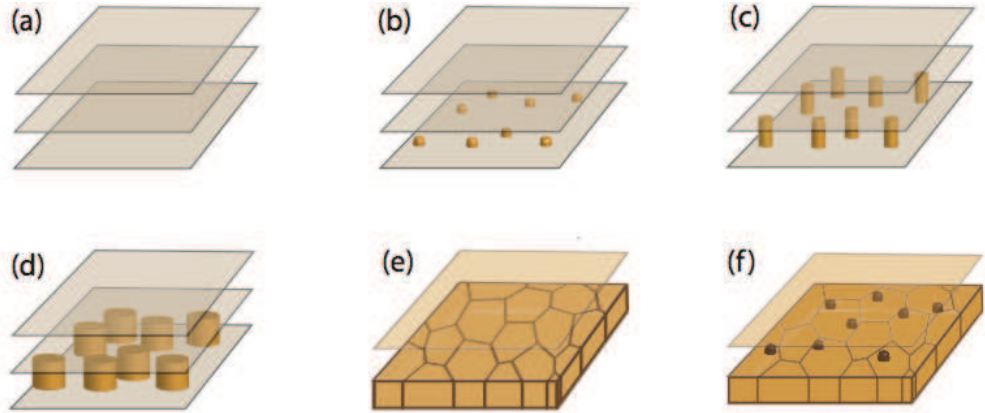}
\vspace{0cm}
%\begin{center}
\caption{Schematic of growth of
sheet nacre, which exhibits layer-by-layer growth.  (a)
Scaffolding of organic matrix sheets is created.
(b) Crystalline aragonite tablets are nucleated at random positions
in the first layer of pre-existing scaffolding
of organic matrix sheets.  (c-e) Aragonite tablets are confined by
organic matrix sheet in the next layer, but grow within the layer until
they reach confluence.  (f)  Aragonite tablets are nucleated
at random positions in the next layer.  With high probability, the
nucleated tablet has the same crystal orientation as that of the
tablet directly below the nucleation site~\protect{\cite{Sch97,Son2003,Car2007}}.
}
%\end{center}
\label{fig:growth_cartoon}
\end{figure}

The model that we examine here assumes that
growth in a given layer is completed before tablets in the succeeding layer are nucleated,
and that the positions
of nucleation sites in all successive layers are nucleated at locations
that are uncorrelated with those in the preceding layers.
The growth rates in the first layer are chosen uniformly at random in the interval $[1-\delta/2, 1+\delta/2]$, and the tablets in each layer grow to confluence. With probability $1-\epsilon$ a tablet has the same growth rate as the tablet below its nucleation site, while with probability $\epsilon$ the tablet is assigned a growth rate chosen uniformly at random from the range $[1-\delta/2, 1+\delta/2]$. 
Appendix A presents the arguments that these choices of the
probability distributions
are natural, given
the geometry of the experimental system.
The version of the model used here and in Ref.~\cite{Met2007b}
is simpler than that examined in
Ref.~\cite{Gil2008}, which
explicitly models columnar nacre by assuming
that growth in a given layer is not completed before tablets
in the succeeding layer are nucleated, so that nucleation sites in successive
layers must be within a certain in-plane distance and are thus correlated;
it is theoretically more attractive because it requires the specification
of one fewer parameter.
Ref.~\cite{Met2007b} presents numerical simulations of the model
and experimental data on the pattern of tablet orientations in red abalone nacre,
and demonstrates that numerical simulations of this
model using parameter values $\epsilon=0.015$ and $\delta=0.25$ yields good
agreement with the experimental
measurements on nacre from red abalone, {\it Haliotis rufescens}.

The next section presents an analytic treatment of the behavior of the
probability distribution function governing the density of tablets of different
orientations as a function of the distance from the prismatic boundary.

\subsection{Growth model used for simulations}
\label{subsec:numerical_model}
This paper reports simulation of a model in which nucleation
sites are placed randomly on a square calculational domain with
open boundary conditions.  In
each layer, the rate of growth of each tablet is chosen at random
from a uniform distribution in the range $[1-\delta/2,1+\delta/2]$.
The tablets nucleate simultaneously and
grow to confluence before
successive layers are nucleated.
The resulting pattern of tablet boundaries in each layer
is a multiplicatively
weighted Voronoi construction~\cite{Oka94,Kob2002a,Mu2006}.
Variations in the in-plane extent of each tablet
within each individual layer are ignored, so that the volume of a given
tablet is the product of its area and the layer separation.
Because faster-growing tablets tend to take up a greater fraction
of the area, and because a tablet will nucleate
with the same velocity as the tablet below it with probability
$1-\epsilon$, when $\epsilon$ is small, there is a tendency
for faster-growing tablets to take up more and more of the total
area (and hence volume) as the growth proceeds.
The ordering is not perfect, however, because some misoriented
tablets are nucleated.

\section{Effective theory for the evolution of probability distribution of tablet orientations}
\label{sec:analytic_theory}
The theory in this section is formulated in terms of crystal orientations, as
opposed to the formulation in terms of growth rates in the previous section.
It is useful to consider
this formulation because the X-PEEM experiments measure
crystal orientations as opposed to growth rates, so comparison of theoretical
predictions to experimental results is facilitated.
Motivated by the ordering of tablet c-axis orientations observed experimentally, our
model posits that variability in the orientations of the c-axes of the tablets
 gives rise to variability in the tablet growth rates, with tablets
 with c-axes aligned perpendicular to the layer plane growing the fastest.

This section presents and analyzes dynamical equations
that govern the evolution of the probability distribution
 describing the number of tablets with different crystal orientations.
These equations can be solved analytically
in the limit that the fraction of misoriented tablets is very small,
which is the parameter regime relevant to the experiments of 
Ref.~\cite{Gil2008}.
We find that the geometry of the dynamical model leads to the result that the
distribution of tablet orientations (or, equivalently, growth rates)
 has an extremely sharp peak whose width
is exponentially small in $\epsilon$, the fraction of misoriented tablets.

The analytic theory presented here for the evolution of tablet orientations
in nacre is closely related to mutation-selection models
 studied in the context of population 
 biology~\cite{Kim65,Cro70,Tur84,Bur96,Cop99,Wax98}.
 The faster growth of tablets of a particular orientation is analogous to
the process of selection in population biology, where species with higher fitness
 reproduce faster than species with lower fitness.
 In the growth model examined in this paper, 
 after all the tablets in a given layer grow to confluence,
 the next layer nucleates.
 When a tablet in the next layer nucleates, one of two things can happen.
 The first, more likely, possibility is that the nucleated tablet has the same orientation
 as the tablet directly below its nucleation site (analogous to inheritance,
 where the descendant has the same fitness as the parent), while
 the second is that the tablet has a randomly chosen orientation (analogous
 to mutation, where the fitness of the descendant differs from that of the
 ancestor by a random amount).
 Selection in the growth model is reflected in the tendency for a larger fraction
 of the area to be filled with tablets with higher growth rates, which increases
 the probability that a given randomly located nucleation point is located
 over a tablet with a higher growth rate.
 
The analytic theory presented here is similar to Ginzburg-Landau 
theories~\cite{Ginz50}
often used in statistical physics~\cite{Goldenfeldbook92}
to describe phase transitions in condensed matter systems, in
that the functional forms follow from symmetry and dimensionality
considerations
and may involve unknown coefficients.
Even if the numerical values of these coefficients are not known,
the analytic theory is very useful for obtaining insight into the
interplay between the various parameters in the problem.
However, additional insight is obtained if the Ginzburg-Landau
parameters can be related to the parameters of the growth
model described above in Sec.~\ref{subsec:numerical_model}, 
which is done in Sec.~\ref{sec:parameters}
using a simple mean-field theory and also
by comparing the predictions of the
analytical model to the results of numerical simulations.

 We define $\phi_\ell(\gamma)d\gamma$ to be the fraction of tablets in layer 
 $\ell$ that are misoriented from the layer normal by angles that are
 between $\gamma$ and $\gamma+d\gamma$.
We assume that
tablets with $\gamma=0$ have the largest rate of growth,
so their share of the area in
layer $\ell+1$ will tend be greater than in layer $\ell$,
thus leading to a larger fraction of tablets with crystal axes
oriented parallel to the layer normal;
the function $w(\gamma)$ governs this tendency.
We define $\chi_\ell(\gamma)d\gamma$ to be the
fraction of the area in layer $\ell$ after its growth is completed
that is oriented in the range of
angles between $\gamma$ and $\gamma+d\gamma$
from the layer normal, and that
\begin{equation}
\chi_{\ell}(\gamma)=\frac{1}{N_{\ell}} w(\gamma)\phi_\ell(\gamma)~,
\end{equation}
where, for each $\ell$, the normalization factor $N_\ell$ is
determined by the normalization of the probability:
\begin{eqnarray}
\label{eq:normalization}
1&=&\int_{0}^{\pi} d\gamma~ \chi_{\ell}(\gamma)\\
\Rightarrow N_{\ell} &=& \int_{0}^{\pi} d\gamma~ w(\gamma)\phi_\ell(\gamma)~.
\end{eqnarray}
The maximum growth rate is at $\gamma=0$, and one expects
$w(\gamma)$ to have a quadratic maximum at $\gamma=0$.
We will scale $w(\gamma)$ so that $w(0)=1$.
We assume that $w(\gamma)$ is a function only of $\gamma$
(in other words, that
it depends only on the degree of misorientation of the c-axis
and not on the orientation of the a and b axes). 
One expects
the dependence of $w(\gamma)$ on $\gamma$ near its maximum
at $\gamma=0$ to be
quadratic, which, if the distribution is not too broad, can be
approximated as a Gaussian,
\begin{equation}
\label{eq:w_equation}
w(\gamma) \propto \left ( 1-\frac{\alpha \gamma^2}{2} \right )
\approx e^{-\alpha\gamma^2/2}~,
\end{equation}
where $\alpha$ is a numerical coefficient.
The tendency for fast-growing tablets to take up an increasing
fraction of the total tablet area is
analogous to the effects of selection in population biology, where
organisms with higher fitness tend to comprise an increasing fraction
of the population.

We then assume that most of
the tablets that nucleate in layers above the first layer have the same
crystal orientations as those of the tablets just below their nucleation sites, but that
there is a small probability $\epsilon$ that a tablet nucleates with
a value of $\gamma$ that is chosen at random
from a normalized probability distribution $f(\gamma)$.
The nucleation of misoriented tablets in the nacre growth model
is analogous to the effects of mutation in a population genetics
model.

The combination of these growth and nucleation terms leads to a
dynamical equation governing
the behavior of $\phi_\ell(\gamma)$:
\begin{eqnarray}
\label{eq:model_def}
\phi_{\ell}(\gamma) &=& \epsilon f(\gamma) +
(1-\epsilon)\chi_\ell(\gamma)\nonumber\\ 
&=& \epsilon f(\gamma) + 
\frac{1}{N_{\ell}}(1-\epsilon) \phi_{\ell-1}(\gamma) w(\gamma)~.
\end{eqnarray}

To complete the definition of the model, one must specify an appropriate
form for $f(\gamma)$, the function describing the distribution of angles of 
misoriented tablet orientations.
We will show below that the behavior depends only on the properties of
$f(\gamma)$ as $\gamma \rightarrow 0$.
In Appendix A
it is found that the generic behavior for $f(\gamma)$ for small
$\gamma$ is
for $f(\gamma)$ to be proportional to $\gamma$ as $\gamma \rightarrow 0$.
This result is intuitively reasonable for this system geometry because
the angular area between $\gamma$ and $\gamma+d\gamma$ is
proportional to $\gamma$ as $\gamma \rightarrow 0$.
At some points
in the analytic treatment below, we will
 choose the specific, mathematically convenient form
$f(\gamma) =\beta \gamma e^{-\beta \gamma^2/2}$.
This choice does not affect any of the results, 
because the behavior of $f(\gamma)$ as
$\gamma \rightarrow 0$ determines the asymptotic behavior.

We characterize the behavior of this model using the methods
of Refs.~\cite{Kin78,Tur84,Cop99}.
First we note that when $\epsilon=0$, so that
the co-orientation of tablets in successive layers
is perfect, this model is easily solved for
any initial distribution, $\phi_1(\gamma)$.
Since increasing
$\ell$ by one multiplies the un-normalized
$\phi_\ell(\gamma)$ by
$w(\gamma)$, it follows immediately that
\begin{equation}
\phi_{\ell}(\gamma) \propto \phi_1(\gamma)(w(\gamma))^{(\ell-1)}~.
\end{equation}
(Note that it is sufficient to compute the un-normalized distribution since the
normalization factor for any given $\ell$
can always be obtained via Eq.~(\ref{eq:normalization}).)
The long-time behavior for any $w(\gamma)$ with a 
single quadratic
maximum at $\gamma=0$ depends only on the curvature in
$w(\gamma)$
near $\gamma=0$.
For the specific choice $w(\gamma)=\exp(-\alpha\gamma^2/2)$,
one finds 
$\phi_{\ell}(\gamma) \propto \phi_1(\gamma) \exp(-\alpha(\ell-1)\gamma^2/2)$.
The width of the distribution
decreases as the square root of the number of layers and becomes
arbitrarily narrow as the number of layers tends to infinity.

Now consider the effects of nonzero but small $\epsilon$, so that a
nonzero fraction of tablets nucleate that are misoriented.
The intuitive picture of the process in this regime
is that the distribution of angles gets narrower unless nucleation
of misoriented tablets
occurs.
The misorientations
prevent the peak from narrowing indefinitely, so 
after many layers $\phi_\ell(\gamma)$ approaches a stationary distribution
that does not change as $\ell$ increases further.
Since a fraction $\epsilon$ of tablets is misoriented at each
layer, one might expect that
the peak in the distribution narrows for $\sim (1/\epsilon)$
layers, so that
this naive argument leads to the
expectation that the eventual width of the distribution should be proportional
to $\epsilon^{1/2}$.
However, it is shown below that
the subtle interplay between the effects of mutation and selection
results in a peak width that can be
exponentially small in $\epsilon$ as $\epsilon \rightarrow 0$.
The high degree of tablet orientation obtained using this mechanism
could be a significant advantage in the biological context.

\subsection{Steady state behavior of the model}
\label{subsec:fixedpoint}
First we find the fixed point behavior for this model, in which
the probability distribution function $\phi_\ell(\gamma)$ approaches
a limit $\phi^*(\gamma)$ that is
independent of $\ell$.
We expect that this fixed point distribution
is reached in the limit $\ell\rightarrow\infty$.

The equation determining the fixed point probability distribution 
$\phi^*(\gamma)$ is
\begin{equation}
\label{eq:phistarequation}
\phi^*(\gamma)=\epsilon f(\gamma)+\frac{1}{N^*}(1-\epsilon)
\phi^*(\gamma)w(\gamma)~,
\end{equation}
with
\begin{equation}
\label{eq:Nstarnormalization}
N^* = \int_{0}^{\pi}d\gamma~\phi^*(\gamma)w(\gamma)~.
\end{equation}
The solution to this equation is
\begin{equation}
\phi^*(\gamma)=\frac{\epsilon f(\gamma)}
{1-\frac{1-\epsilon}{N^*}w(\gamma)}~,
\end{equation}
with
\begin{equation}
\label{eq:nstarselfconsistency}
N^*=\int_{0}^{\pi}d\gamma \epsilon f(\gamma) w(\gamma)
\sum_{m=0}^{\infty}
\left [ \frac{1-\epsilon}{N^*} w(\gamma)\right ]^m~.
\end{equation}

%Using the definition of $N^*$ from Eq.~\ref{eq:Nstarnormalization} yields
%\begin{eqnarray}
%\label{eq:selfconsistency1}
%N^* &=& \int_{-\pi/2}^{\pi/2}w(\gamma)\phi^*(\gamma)d\gamma\nonumber\\
%&=& \epsilon \int_{-\pi/2}^{\pi/2} d\gamma~f(\gamma)w(\gamma)
%\sum_{m=0}^{\infty}\left [ 
%\frac{1-\epsilon}{N^*}w(\gamma)
%\right ]^m~.
%\end{eqnarray}

We now define
\begin{equation}
\label{eq:vdefinition}
v=\frac{1-\epsilon}{N^*}~,
\end{equation}
\begin{equation}
\label{eq:Idefinition}
I_m = \int_{0}^{\pi} d\gamma(w(\gamma))^{m+1}f(\gamma)~,
\end{equation}
and rewrite Eq.~(\ref{eq:nstarselfconsistency}) as
\begin{equation}
\label{eq:compactselfconsistency}
\frac{1-\epsilon}{v} = \epsilon \sum_{m=0}^\infty v^m I_m~.
\end{equation}

A useful explicit form for the area distributions can be obtained
in the parameter regime in which $\epsilon$, the fraction of misaligned tablets, is small.
When $\epsilon \ll 1$, the right hand side of Eq.~(\ref{eq:compactselfconsistency}) can
be of order unity only if $v^m I_m$ decays slowly for large arguments.
Because of its definition, $v > 0$, and
it will be seen below that normalization of the probability implies that
$v \le 1$.
Therefore,
when $m\gg 1$, because $w(\gamma)$ has a single maximum at
$\gamma=0$, the integrand in $I_m$ from Eq.~(\ref{eq:Idefinition})
is very sharply peaked near $\gamma=0$.
Therefore, the integration interval can be extended to $[0,\infty]$,
one can assume that $w(\gamma)$ is a Gaussian,
$w(\gamma)=\exp(-\alpha\gamma^2/2)$,
and only the behavior of $f(\gamma)$ for small $\gamma$ is relevant.
If the orientations of the misoriented tablets are chosen uniformly at random
in three-dimensional space, then, as discussed in~\ref{appendix:dimensionality},
$f(\gamma)$ is proportional to $\gamma$ as
$\gamma \rightarrow 0$, and for small argument $f(\gamma)=\beta\gamma$
(with $\beta$ a constant of order unity),
so that
$I_m=\beta\int_{0}^\infty d\gamma ~\gamma ~\exp(-\alpha (m+1)\gamma^2/2)=\beta/(\alpha(m+1))$.
Therefore, when $\epsilon \ll 1$,
\begin{eqnarray}
\label{eq:v_equation}
\frac{1-\epsilon}{v} &=& \epsilon \frac{\beta}{\alpha}\sum_{m=0}^\infty \frac{v^m}{m+1}~.
\end{eqnarray}
(Eq.~\ref{eq:v_equation} makes it particularly clear that 
a solution is possible only if $v \le 1$.)
Using the identity~\cite{lewinbook81}
\begin{equation}
\sum_{k=1}^\infty \frac{x^k}{k}=-\ln(1-x)~,
\end{equation}
%(see 
%Weisstein, Eric W. "Polylogarithm." From MathWorld--A Wolfram Web Resource. http://mathworld.wolfram.com/Polylogarithm.html; 
%Lewin, L. Polylogarithms and Associated Functions. North-Holland, New York. 1981)
one finds
\begin{equation}
\frac{1-\epsilon}{\epsilon}\frac{\alpha}{\beta}=-\ln(1-v)~,
\end{equation}
so that
\begin{equation}
N^*=\frac{1-\epsilon}{v}\approx (1-\epsilon)
\left (1+\epsilon \left [ -\frac{\alpha(1-\epsilon)}{\beta\epsilon}\right ] \right )~.
\end{equation}
Therefore, $\phi^*(\gamma)$ can be written
\begin{equation}
\label{eq:phi_star_solution}
\phi^*(\gamma) = \frac{\epsilon f(\gamma)}
{1-\left (1-\exp \left [-\frac{\alpha(1-\epsilon)}{\beta\epsilon} \right ]
\exp[-\alpha \gamma^2/2] \right )}~.
\end{equation}
The width of the probability distribution $\phi^*(\gamma)$ in $\gamma$,
estimated by finding the value of $\gamma$ at which
$\phi^*(\gamma)=\phi^*(0)/2$, is 
\begin{equation}
\gamma_{1/2}=\sqrt{\frac{2}{\alpha}}e^{-(1-\epsilon)\alpha/(2\beta\epsilon)}~,
\end{equation}
which is extremely small when $\epsilon$ is small.

\subsection{Dynamics of the analytic model}
\label{subsec:dynamics}
The dynamics of the model defined in Eq.~(\ref{eq:model_def}) can of course
be obtained numerically.
Analytic insight can also be obtained,
following Ref.~\cite{Cop99}, by
writing Eq.~(\ref{eq:phistarequation}) for the fixed point function as
\begin{equation}
\phi^*(\gamma)=\epsilon f(\gamma)\sum_{m=0}^\infty \left (
\frac{1-\epsilon}{N^*}w(\gamma) \right )^m~,
\end{equation}
and comparing this expression to the solution to Eq.~(\ref{eq:model_def}),
which can be written
\begin{equation}
\phi_\ell(\gamma)=\epsilon f(\gamma) \left [\sum_{m=1}^{\ell-1}
\prod_{n=1}^{m-1} \left ( \frac{1-\epsilon}{N_n} w(\gamma) \right )
+\prod_{n=1}^\ell \left ( \frac{1-\epsilon}{N_n} w(\gamma) \right )\phi_1(\gamma)
\right ]~.
\end{equation}
If one assumes that the term proportional  to $\phi_1(\gamma)$ is negligible
and that the normalizations $N_n$ can be approximated as being the
same, $N_n=\mathcal{N}$, independent of $n$, then
one obtains an expression for $\phi_\ell(\gamma)$:
\begin{equation}
\label{eq:dynamic_approx}
\phi_\ell(\gamma) \approx \epsilon f(\gamma)
 \sum_{m=1}^{\ell-1} \left (\frac{1-\epsilon}{\mathcal{N}} w(\gamma) \right )^m~.
\end{equation} 
This expression is the sum of
contributions that can be interpreted as describing the contribution of
the population that has undergone a given number of selection
events since the last mutation,
and Ref.~\cite{Cop99} shows that it agrees well with numerical solutions
of Eq.~(\ref{eq:model_def}).

\section{Relating the parameters in the effective theory of Sec.~\protect\ref{sec:analytic_theory}
to the parameters in the growth model of Subsec.~\protect\ref{subsec:numerical_model}.}
\label{sec:parameters}
Sec.~\ref{sec:analytic_theory} presents an analysis of a theory in which
one writes an equation for $\phi_\ell(\gamma)$, the fraction of the
area in layer $\ell$ in which the tablet orientation angle is $\gamma$.
The model analyzed there contains three parameters.  The first, $\alpha$,
defined in Eq.~\ref{eq:w_equation}, governs the degree
to which 
tablets with small values of $\gamma$ grow faster than tablets with
larger values of $\gamma$.
The second, $\epsilon$,
is the fraction of nucleation sites that have tablets with 
a random orientation instead of the same orientation as the tablet below.
The third parameter,
$\beta$, specifies the width of the distribution governing the
distribution of angles of the misoriented tablets.
In contrast, the original growth model described in
Subsec.~\ref{subsec:numerical_model} has two parameters,
$\epsilon$, the fraction of misoriented tablets, and
$\delta$, which governs the range of in-plane growth speeds of the misoriented tablets.
This section describes the relationships between the two sets of parameters.

The value of $\epsilon$ is the same in
the analytic theory for orientations
as in the growth model (hence the use of
the same symbol).
The
parameters $\beta$ and $\delta$ are related in a straightforward
fashion, as discussed below in subsection~\ref{subsec:beta_delta}.
Most of this section will focus on the
third parameter in the effective theory, $\alpha$, 
which we show depends on the geometry
of the nucleation sites in the growth model
in subsection~\ref{subsec:geometry_dependence_synchronous}.
We will present a simple
mean-field theory for estimating the value of $\alpha$
for uncorrelated and random nucleation site locations
which
yields the correct order of magnitude for the value
in subsection~\ref{subsec:mean_field_theory}.
We compare the mean field predictions with the
results of simulations of
the growth model in subsection~\ref{subsec:comparison}.

\subsection{Analytic theory for model formulated in terms of tablet growth rates}
\label{subsec:analytic_rates}
In the growth model that is simulated numerically,
the tablet orientation angles are not considered explicitly because
the calculation is formulated using tablet growth rates, and
the simulation is performed by choosing an initial configuration of
nucleation sites with a
distribution of growth rates, and then allowing the tablets to grow
from these nucleation sites until they reach confluence.
We choose to define a new variable
$x$ that ranges between $0$ and $1$, and
define $\mathcal{P}_\ell(x)$, the probability distribution describing 
the relative frequencies of tablets in layer $\ell$
with different values of $x$~\cite{Kin78}.
It is natural to interpret
$x$ as $v/v_{max}$, where $v$
is the in-plane tablet growth speed and $v_{max}$ is the tablet
growth velocity at the orientation where this velocity is maximum.

As discussed above and
in \ref{appendix:dimensionality}, when the model is formulated
in terms of orientation angles, the ``fitness"
function specifying the changes in the fractions of the area covered by tablets with 
different
angles of misorientation between successive layers
is expected to have a quadratic
maximum at $\gamma=0$.
Because the tablet growth velocity itself 
depends quadratically
on $\gamma$ near $\gamma=0$ (again, because misorientations
by $\gamma$ and $-\gamma$ are equivalent and so yield
the same growth velocity),
the ``fitness" function that specifies the change in relative area
of the different values of $x$, $\tilde{w}(x)$,
depends linearly on $x$ near
$x=1$.
Recalling that the probability distributions in the model are
normalized, so that the overall scale of $\tilde{w}(x)$ is arbitrary,
and that the behavior is dominated by the behavior near $x=1$,
two ways of parameterizing this dependence are to (1) fix
the value of $\tilde{w}(1)=1$ and specify
the slope $\tilde{w}^\prime$, or (2) to
write $\tilde{w}(x)=x^\xi$ and specify $\xi$.
These two forms are equivalent near $x=1$, with $\tilde{w}^\prime(x)=\xi$;
we will choose to use the power-law form $\tilde{w}(x)=x^\xi$.

To obtain additional insight into the relationship between the
two formulations of the model, we reformulate the analytic theory
in terms of the variable $x$ (this formulation is very similar
to that in Ref.~\cite{Kin78}).
Assuming that tablets with normalized growth rate $x$
grow to have an area that is proportional to $x^\xi$ for some $\xi$,
and assuming
that there the probability of nucleation of a misoriented tablet
is $\epsilon$, the probability
distribution for the tablet growth velocities in layer $\ell$,
$\mathcal{P}_\ell$, obeys 
\begin{equation}
\mathcal{P}_{\ell+1}(x) = \epsilon g(x)
+(1-\epsilon) (1/\mathcal{N}_\ell)x^\xi \mathcal{P}_\ell(x)~,
\end{equation}
where $g(x)$ is the probability distribution for the misoriented
tablets and
$\mathcal{N}_\ell=\int_0^1 dx x^\xi \mathcal{P}_\ell(x)$.
Using methods precisely analogous to those in the previous section,
we write the equation for
 fixed point reached at large $\ell$, $\mathcal{P}^*(x)$:
 \begin{eqnarray}
 \mathcal{P}^*(x) &=& \frac{\epsilon g(x)}{1-\frac{(1-\epsilon)}{\mathcal{N}^*}x^\xi}
 \nonumber\\
 &=& \epsilon g(x)\sum_{m=0}^\infty \left ( 
 \frac{(1-\epsilon)}{\mathcal{N}^*}x^\xi \right )^m
 ~,
 \end{eqnarray}
where $\mathcal{N}^*$ obeys
\begin{eqnarray}
\mathcal{N}^*&=& \int_0^1 dx \mathcal{P}^*(x)\nonumber\\
&=& \int_0^1 dx \epsilon g(x)\sum_{m=0}^\infty \left (
\frac{(1-\epsilon)}{\mathcal{N}^*}x^\xi \right ) ^m~.
\end{eqnarray}
When $\epsilon$ is small, then the integrals are dominated by the region
where $x$ is close to unity, and
the $x$-dependence of $g(x)$ can be neglected,
yielding
\begin{equation}
\mathcal{N}^*= \epsilon g(1)\sum_{m=0}^\infty
\left (\frac{1-\epsilon}{\mathcal{N^*}} \right )^m \frac{1}{1+\xi m}~.
\end{equation}
When $\epsilon$ is small, so that $v$ is close to $1$, this
sum is dominated by the region in which $\xi m \gg 1$, and,
using $\sum_{k=1}^\infty x^k/k=-\ln(1-x)$, one obtains
\begin{eqnarray}
\mathcal{N}^* &=& 
-\frac{\epsilon g(1)}{\xi}
\ln\left (1-\frac{1-\epsilon}{\mathcal{N^*}} \right )~.
\end{eqnarray}
Recalling that $\mathcal{N}^*$ is extremely close to $1-\epsilon$, 
one obtains
\begin{equation}
1-\epsilon = -\frac{\epsilon g(1)}{\xi}
\ln\left (1-\frac{1-\epsilon}{\mathcal{N^*}} \right )~,
\end{equation}
so that 
\begin{equation}
\mathcal{N}^*=(1-\epsilon)
\left ( 1+\exp -\left ( \frac{(1-\epsilon)\xi}{\epsilon g(1)} \right ) \right )~
\end{equation}
and
\begin{equation}
\label{eq:mathcalPequation}
\mathcal{P}^*(x) = \frac{\epsilon g(x)}
{1-\left (1-\exp\left [ -\frac{(1-\epsilon)\xi}{\epsilon g(1)}\right ] \right )x^\xi} ~.
\end{equation}
This form is consistent with Eq.~(\ref{eq:phi_star_solution}).
Comparing the forms, one can see that the parameter $\alpha$
in the effective model enters into Eq.~(\ref{eq:phi_star_solution})
in a way similar to that in which $\xi$ enters into Eq.~(\ref{eq:mathcalPequation}),
and that 
$\beta$ enters 
into Eq.~(\ref{eq:phi_star_solution})
in a way similar to that in which $g(1)$ enters into Eq.~(\ref{eq:mathcalPequation}).

\subsection{Relationship between parameters $\beta$ and $\delta$.}
\label{subsec:beta_delta}
The parameter $\beta$ in the analytic model and the parameter $\delta$
in the growth model each describe the probability distribution governing
the misoriented tablets.
As seen above,
because of the action of the ``selection" in the model,
the key property of these distributions is their behavior
near $\gamma=0$ and $x=1$.
Therefore, we work out the relationship between the two formulations
of the model in this region.

For the growth model, the distribution of the in-plane growth speed $v$
is chosen to be uniform in the range $[1-\delta/2,1+\delta/2]$, so the
probability density in terms of the variable $v$ is $1/\delta$.
To convert to $x$, the normalized growth speed, we note that
the maximum growth velocity is $1+\delta/2$, so that the probability
density for $x$ is $(1/\delta)(dv/dx)=(1+\delta/2)/\delta$.
For the analytic formulation of the theory, choosing this distribution to be
$f(\gamma)=\beta \gamma e^{-\beta \gamma^2/2}$
(the coefficient is determined by normalizing
$\int_0^{\pi/2}f(\gamma)d\gamma = 1$, and assuming
that $\beta$ is large enough that setting the upper limit of the integration
at $\infty$ introduces negligible error),
the probability density near $\gamma=0$ is $f(\gamma) \rightarrow \beta\gamma$.
Near $\gamma=0$, the variables $\gamma$ and $x$ are related
by $x=1-A\gamma^2$, for some constant $A$, so near $\gamma=0$ we can write
\begin{eqnarray}
d\gamma f(\gamma) &\approx& dx (1+\delta/2)\delta\nonumber\\
dx (d\gamma/dx)(\beta\gamma) &\approx& dx (1+\delta/2)/\delta\nonumber\\
(\beta\gamma)/(2A\gamma)&\approx& (1+\delta/2)/\delta~,
\end{eqnarray}
leading to the identification $\beta/(2A)=(1+\delta/2)/\delta$.
This relationship between the probability distributions of the orientation
angles and the growth
speeds involves not just $\beta$ and $\delta$ but also a new constant $A$,
that specifies the relationship between the change in orientation angle and
the change in growth speed.
However, this constant
$A$ also enters into the ``fitness" or ``selection"
term that is discussed
in the next subsection in precisely analogous fashion,
so that the relationship between ratios $\alpha/\beta$
and $\xi/g(1)$ in Eqs.~(\ref{eq:phi_star_solution})
and Eq.~(\ref{eq:mathcalPequation}) does not depend on $A$.

\subsection{Dependence of $\alpha$ on geometry}
\label{subsec:geometry_dependence_synchronous}
This subsection discusses how the parameter $\alpha$ in Eq.~\ref{eq:w_equation}
is related to the growth model as described in Subsec~\ref{subsec:numerical_model}.
It can be seen that $\alpha$ does not depend on $\epsilon$ and $\delta$
by noting that in
the limit of $\epsilon\rightarrow 0$,
when no misoriented tablets are nucleated, both $\epsilon$
and $\delta$ drop out of the model altogether, while
the selection term in which $\alpha$ appears still operates.
We will see that the value of $\alpha$
 is essentially determined by geometrical considerations only.
 
The value of $\alpha$ in the analytic model depends on
the geometric arrangement of nucleation sites in the growth model.
This is because
the areas of the tablets after growth is completed depend on the
differences between growth rates of neighboring tablets as opposed
to depending on individual growth rates themselves;
two neighboring tablets growing at the same speed will not lose
area to each other, no matter what that rate happens to be.
Changes in the fractions of area covered by tablets with different
growth rates arise only when
fast-growing tablets are next to slow-growing tablets. 
In fact, if
one considers a one-dimensional case of $N$ tablets in which the
growth rate of tablet $j$ is $g_0+cj$, where $g+0$ and $c$ are
constants, it is easy to see that there is a set of locations for the 
nucleation sites such that
the size of every tablet stays the same except for the tablets
at $j=1$ and $j=N$.

Here we consider the specific case of a collection of uncorrelated,
randomly chosen nucleation sites.
For this situation, neighboring
tablets are likely to have different growth
speeds, so that the differential growth rates will lead to fast-growing tablets 
increasing their share of the area.
We first present a simple mean field estimate for the resulting
value of $\alpha$, and then present numerical computations
that can be used to extract its value.

\subsubsection{Mean field theory for $\alpha$ for random nucleation sites}
\label{subsec:mean_field_theory}
In this subsection we construct a simple mean-field theory for estimating $\alpha$.
In the next subsection we will compare the mean-field theory results
with the results obtained using
numerical simulations.

In subsection~\ref{subsec:analytic_rates}
we showed that without loss of generality, one
can choose the form for the selection
term in the growth model for the relative area
covered by tablets with normalized growth speed $x$ to
be $x^\xi$.
Here we present a
simple mean-field theory for calculating $\xi$ using arguments very
similar to those
typically used in statistical physics~\cite{Chandlerbook87}:
We assume that when a tablet grows, it does so in the presence
of neighbors with $mean$ properties.
 and take account explicitly
of the fact that faster-growing tablets will tend to reach confluence with their
neighbors sooner than slower-growing ones.
One assumes that
when a given tablet
grows, it does so in the presence of neighbors that have $average$ properties.

Given that a tablet has normalized growth speed
 $x=v/v_{max}$ and that the mean normalized growth speed is
 $\bar{x}=\bar{v}/v_{max}$, then we wish to calculate the
 area covered by the tablet after it has grown to confluence with its
 neighbors.
For simplicity, we will consider nucleation sites that are placed on a regular
lattice (or alternatively, consider the scaling of the behavior without
worrying about numerical coefficients).
For two nucleation sites separated by $d$ with normalized
growth rates $x_1$ and $\bar{x}$, the time for the tablets to grow
to confluence is proportional to $d/(x_1+\bar{x})$,
so the area of the tablet in question should scale as
${x_1}^2 /(x_1+\bar{x})^2$.
Therefore, the ratio of the area taken up by the
tablet we are considering to that of the ``mean-field" tablet (which we assume
has area that scales as $\bar{x}^2/(x_1+\bar{x})^2$) is
${x_1}^2/\bar{x}^2$.
Approximating $\bar{x}$ as constant (which is reasonable because
the probability distribution for $x$ is asymmetric, monotonically
decreasing from $x=1$, with a mean very near 1) yields the result that
the area growth for a normalized velocity $x_1$ is proportional
to $x_1^2$, so that $\xi=2$.

A modification of this mean field theory, that yields a value for $\xi$ that
agrees better with the results of the numerical
simulations below, is obtained by noting that while the area of the tablet under consideration
is proportional to ${x_1}^2 /(x_1+\bar{x})^2$, the area of the mean-field neighbors
varies little when $x_1$ changes
because their interactions with their other neighbors are not affected.
Assuming that the change in areas of the ``mean-field'' tablets
 can be neglected altogether
yields the value $\xi=1$.
This result can be seen in several ways:
first, the mean-field argument given just above assumes that increasing $x_1$ leads
to a decrease in the ``mean-field" tablet
area that scales the same way as the increase in the area of the ``non-mean-field" tablet,
so ignoring the former contribution leads to a result that is the square root of the one
obtained previously.  Alternatively, one can argue as follows:
since the area growth of the ``non-mean-field" tablet
growing with normalized velocity $x_1$ is proportional
to $x_1^2/(x_1+\bar{x})^2$, we can
write $x_1=1-\rho_1$, $\bar{x}=1-\bar{\rho}$, assume that $\rho_1$
and $\bar{\rho}$ are each $\ll 1$, and fix the proportionality constant by
requiring the relative
area to be unity when $x_1=1$. When $\bar{\rho}$ is close to
unity, this procedure also yields the result $\xi=1$.

\subsubsection{Numerical evaluation of parameters in effective theory for synchronous nucleation and random nucleation geometry}
\label{subsec:comparison}

For the simulations presented in this section,
the initial distribution of tablet growth velocities
tablets are chosen to be uniform in the range $[1-\delta/2,1+\delta/2]$,
or, equivalently,
$g(x)=1/\delta$ for $x$ in the range $[(1-\delta/2)/(1+\delta/2),1]$.
Because the focus is on determining the value of $\alpha$ in the
selection term of the effective model, the parameter
$\epsilon$ is set to zero in the simulations in this section.

The comparison between mean field theory and simulation is
done by comparing $x_{1/2}(\ell)$, the 
median growth speed found in the simulation at layer $\ell$,
to the median of the distribution found using the analytic
mean-field theory,
which satisfies
\begin{equation}
\int_0^{x_{1/2}(\ell)} \mathcal{P}_\ell(x)=1/2~.
\end{equation}

When $\epsilon=0$, the dynamical equation for $\mathcal{P}_\ell(x)$ is
\begin{equation}
\mathcal{P}_{\ell+1}(x)\propto x^\xi\mathcal{P}_{\ell}(x)~,
\end{equation}
which has the solution
\begin{equation}
\mathcal{P}_{\ell}(x)=x^{\xi\ell}\mathcal{P}_0(x)~.
\end{equation}
When $\ell$ is large, the $x$-dependence of $\mathcal{P}_0(x)$
can be neglected, and one obtains $\mathcal{P}_\ell(x)\propto x^{\xi\ell}$,
yielding a normalized distribution
$\mathcal{P}_\ell(x)=(\xi\ell+1)x^{\xi\ell}$.
 The median value of this distribution, $x_{1/2}(\ell)$, is
 \begin{eqnarray}
 \label{eq:xi_equation}
 x_{1/2}(\ell) &=& \left (\frac{1}{2} \right ) ^{\frac{1}{\xi\ell+1}} \nonumber \\
 &\approx& 1 -\frac{\ln 2}{\xi\ell+1}~.
 \end{eqnarray}
 Therefore, when $\ell$ is large, the slope of the plot of
 the median value $x_{1/2}(\ell)$ versus $1/\ell$
approaches $\ln(2)/\xi$, providing a method for extracting
the value of $\xi$ from numerical simulations. 

The procedure used for the numerical simulations is very similar to
that used for those presented in Ref.~\cite{Gil2008}, but specialized
to the limit in which successive layers nucleate much more slowly
than individual layers grow to completion.
Each layer in the simulation is a square of dimension $400 \times 400$
(the length unit is arbitrary, and is adjusted to agree with measured tablet
 widths when numerical results are compared with experimental data
 in Ref.~\cite{Met2007b,Gil2008}) 
 with nucleation sites whose sites are
 independently and randomly chosen from a probability distribution 
 with uniform spatial density.
 Open boundary conditions were employed.
 The growth rates in the initial layer are chosen uniformly at random in the interval 
 $[1-\delta/2, 1+\delta/2]$ (resulting in normalized growth rates
 in the interval $[(1-\delta/2)/(1+\delta/2),1]$). 
 In these simulations, the parameter $\epsilon$ is set to zero
 (no misoriented tablets
 are nucleated).

\subsubsection{Numerical results}
\label{sec:results}
As discussed in Sec.~\ref{subsec:comparison},
the value of $\xi$ that determines the strength of the ``selection''
can be extracted by computing
the median value of the in-plane tablet growth velocity,
$x_{med}(\ell)$ as a function
of layer number $\ell$, when $\epsilon=0$.
Eq.~(\ref{eq:xi_equation}) predicts that
the quantity $1/(1-x_{med}(\ell))$ depends
linearly on $\ell$, and the slope of this dependence is $\xi/\ln 2$.
For the uniform distribution of growth rates used in the numerical
simulations, the initial value of this quantity,
$1/(1-x_{med}(1))$,  is $1+2/\delta$.

Figure~2 is a plot of the quantity
$1/(1-x_{med}(\ell))$ as a function of layer number $\ell$
for different values of $\delta$.
The graph also shows the dependence expected for
$\delta=0.3$ using
the two different mean-field results obtained above,
$\xi=2$ and $\xi=1$.
There is substantial scatter in the numerical
data, but it appears that the behavior is consistent with
the value $\xi=1$ but not with the value $\xi=2$.
\begin{figure}[ht]
\begin{center}
\vspace{0.5cm}
\hspace{0cm}
\includegraphics[height=6cm]{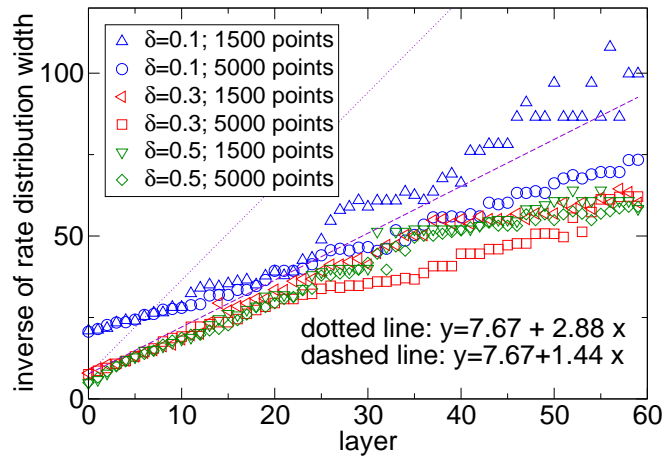}
\vspace{0cm}
\caption{Plots of $1/(1-x_{med}(\ell))$ versus layer number $\ell$, obtained
by numerical simulation of the model,
where $x_{med}(\ell) $ is the median value of normalized
growth speed $x$ in layer $\ell$.
The number of nucleation points per layer and the value
of $\delta$ for each run are shown in the legend.
For the uniform distribution of growth rates used in the numerical
simulations, the initial value of the ordinate,
$1/(1-x_{med}(1))$,  is $1+2/\delta$.
The mean field theories described in the text yield a linear
dependence for this quantity, one with
 slope $2/\ln 2 \approx 2.88$,
corresponding to the value $\xi=2$ (a dotted line with this slope is
shown on the plot), and the other with slope
$1/\ln 2 \approx 1.44$ (dashed line on the plot).
The numerical results exhibit considerable scatter, but
appear to be consistent with the value of $\xi=1$ and
not the value of $\xi=2$.
}
\end{center}
\label{fig:alpha_plot}
\end{figure}

\section{Discussion}
\label{sec:discussion}
In this paper we have detailed investigations of a model
for the development of orientational order of the aragonite tablets
in nacre.
The model is investigated both by direct numerical simulation
and by analysis of the evolution of the probability distribution
governing tablet orientations that is analogous to mutation-selection
theories studied in population biology.
The relationships between the parameters in the version
of the model used for the numerical simulations and those in the
version used for the analytic work are investigated.

The model simulated in this paper can be extended in many ways
to model nacre growth more realistically.
In Ref.~\cite{Gil2008} the model is modified to take the columnar
growth of abalone nacre into account, by restricting each nucleation
site to be within a given distance from a nucleation site in the layer below,
as opposed to the model used in this paper, in which the nucleation
sites in each layer are chosen independently and randomly in
the simulational domain.
Compared to the version in Ref.~\cite{Gil2008},
the model examined here has the advantage that it has one
less parameter.
Because the analytic theory is constructed using very general
symmetry considerations, it is clear that
the qualitative behavior of a model implementing columnar
growth is the same as for the model studied here.
Similarly, the growth model can be generalized in other ways
(for instance,
by implementing more realistic microscopic growth rates with
explicit modeling of three-dimensional growth or by considering
asynchronous
nucleation)
will still
yield the same analytic model, though with possibly different
values for the model parameters.

The relationships between the parameters in the analytic theory
and in the growth model presented in Subsec.~\ref{subsec:numerical_model}
were investigated using a mean field theory and by numerical simulation.
Two mean field theories were constructed
to
 estimate the parameter
 describing the strength of the tendency for
faster-growing tablets to take up more area than slower-growing ones.
Numerical simulations of the growth model yield a result for this
parameter that agree reasonably well with one of the mean field theories.
Development of a more sophisticated analytic theory for calculating the
value for this parameter 
is an interesting open problem.

\section{Conclusions}
\label{sec:conclusions}
This paper investigates theoretically a model for the development
of orientational order of the aragonite tablets in nacre,
or mother-of-pearl.
Motivated by experiments demonstrating the this ordering
develops over many tens of layers, the model assumes that
the tablet growth rates depend on the orientation of their
c-axes, with
tablets with c-axis orientation normal to the layer plane
growing the fastest.
This model is closely related to mutation-selection models used
in population biology.

A combination of analytic and numerical results were applied
to analyze the model.
It is shown that the model yields a
degree of ordering that is extremely good when the parameter
$\epsilon$, that governs the fraction
of tablets with nucleate with misoriented c-axes, is small.

\section{Acknowledgments}
\label{sec:acknowledgments}
We gratefully acknowledge for this research provided by NSF-DMR and by DOE.
S.N.C. acknowledges the hospitality of
the Aspen Center for Physics,
where some of the work was performed.

\appendix
\renewcommand\thesection{Appendix \Alph{section}}
\section{Form of functions used in analytic model}
\label{appendix:dimensionality}
This appendix discusses the forms of the functions $w(\gamma)$ and
$f(\gamma)$ that are used in Sec.~\ref{sec:analytic_theory} to specify the
model for the evolution of the probability distribution of tablet orientations.
The function $w(\gamma)$ specifies the tendency of
the area covered by faster-growing tablets to increase,
while $f(\gamma)$ specifies the probability distribution function for the
angles of the misoriented tablets.
We assume that the growth rate is isotropic in the a-b plane, so
that the tablet growth velocity depends only on $\gamma$,
the angle between
the c-axis orientation of the tablet and the layer normal.

\begin{figure}[ht]
\label{fig:gamma_geometry}
\vspace{0cm}
\hspace{0cm}
\includegraphics[height=10cm]{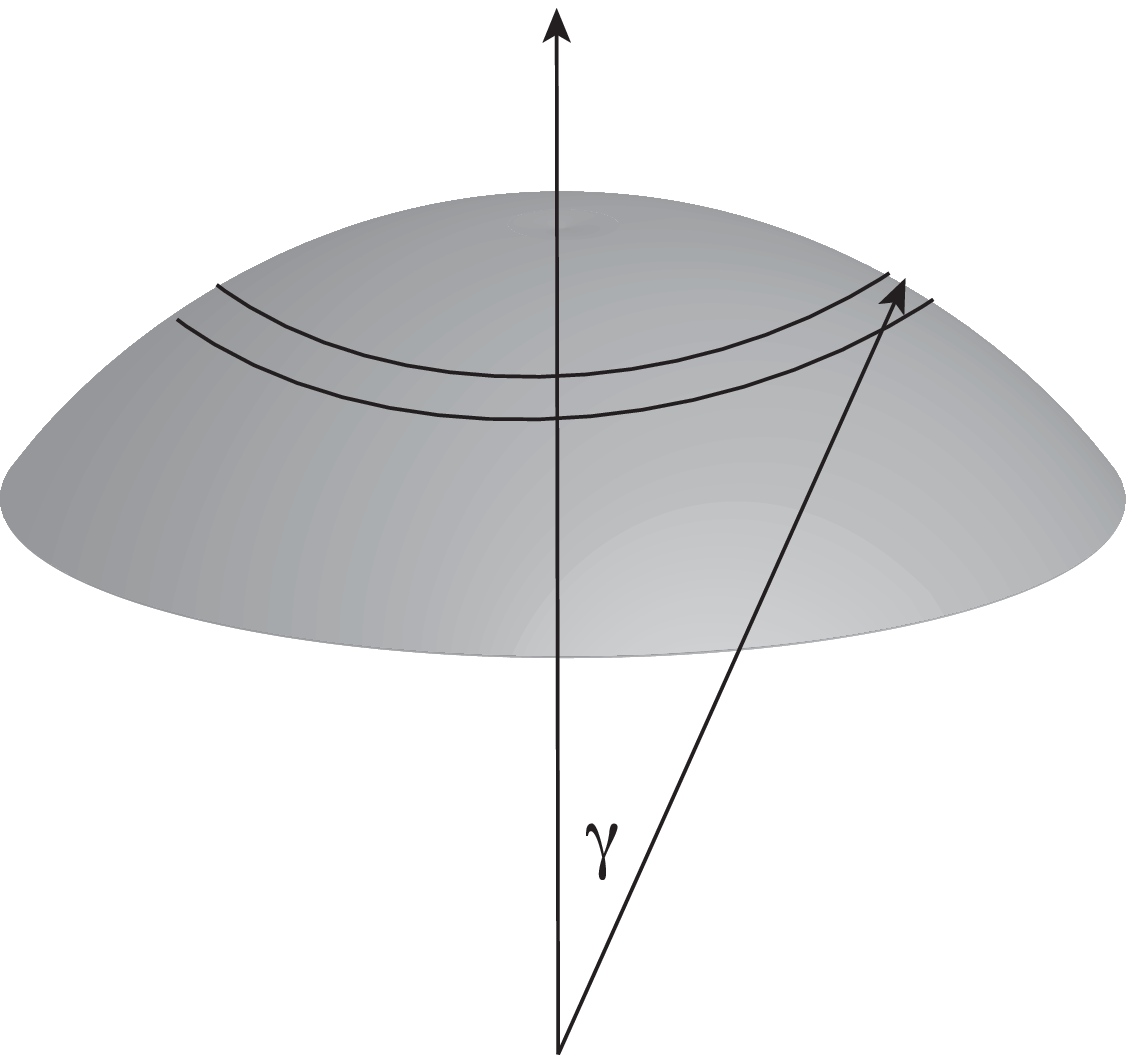}
\vspace{-.1cm}
\begin{center}
\caption{Schematic of geometry, illustrating that
assuming that the c-axes of misoriented tablets are equally likely
to have any orientation, then the function $f(\gamma)$
introduced in Sec.~\ref{sec:analytic_theory} is proportional to $\gamma$
as $\gamma \rightarrow 0$.
}
\end{center}
\end{figure}
For the function $w(\gamma)$,
our assumption that the growth speed has no azimuthal dependence implies
that the growth speed of a tablet with crystal axis
misoriented from the layer normal by an angle $\gamma$ is
a function of $\cos(\gamma)$ and has a quadratic maximum at $\gamma=0$.
When the parameter $\epsilon$ governing the
fraction of misoriented tablets is small, the dynamics is controlled
by the behavior for small $\gamma$.
As discussed in the main text, it is often convenient to choose $w(\gamma)$
to be a Gaussian, which makes calculation of its dynamical effects
particularly simple.

Now we discuss the function $f(\gamma)$ that
describes the distribution of misoriented tablets.
If one makes the simplest assumption
that the orientations of these misoriented tablets take
on each angle with equal probability, then the probability
distribution for $\gamma$,
the angle between the tablet c-axis orientations and a
fixed external axis (here, the direction of the layer normal)
is proportional to $\sin(\gamma)$, since
the area of a spherical surface between angles $\gamma$ and
$\gamma+d\gamma$ is proportional to $\sin(\gamma)d\gamma$,
or $d\cos(\gamma)$.
The relevant geometry is shown in Fig.~3.

As discussed in the main text, the model can be formulated
entirely in terms of tablet growth rates, and we use such a form
in our numerical simulations.
We define the variable $x$ to be the normalized tablet
growth rate, with $x=1$ the maximum value.
Because (1) we assume that the  tablet growth rate exhibits
a quadratic
maximum at $\gamma=0$, and (2) the area between
the angles $\gamma$ and $\gamma+d\gamma$ is proportional
to $\gamma$ as $\gamma \rightarrow 0$,
in terms of $x$, the growth rate is maximum
at $x=1$ and depends linearly on $x$ near $x=1$.
Similarly, the function describing the relative change in
area taken up by tablets with c-axis orientation $\gamma$,
$w(\gamma)$,
has a quadratic maximum as a function of $\gamma$, so
the analogous function expressed in terms of the growth rate
will depend
linearly on $x$ near $x=1$.
 
\section*{References}
\bibliographystyle{unsrt}
\bibliography{biomineralization2}

\end{document}